\documentstyle[aps,preprint]{revtex}

\begin{document}
\tighten
\renewcommand{\arraystretch}{1.5}
\newcommand{\be}{\begin{equation}}
\newcommand{\ee}{\end{equation}}
\newcommand{\bea}{\begin{eqnarray}}
\newcommand{\eea}{\end{eqnarray}}
\def\Tr{\mathop{\rm Tr}\nolimits}
\def\mapright#1{\smash{\mathop{\longrightarrow}\limits^{#1}}}
\def\mapdown#1{\big\downarrow \rlap{$\vcenter
  {\hbox{$\scriptstyle#1$}}$}}
\def\su#1{{\rm SU}(#1)}
\def\so#1{{\rm SO}(#1)}
\def\sp#1{{\rm Sp}(#1)}
\def\u#1{{\rm U}(#1)}
\def\o#1{{\rm O}(#1)}
\def\p#1{{\pi_{#1}}} 
\def\Z{{\bf Z}}
\def\R{{\bf R}}
\def\M{{\cal M}}
\def\L{{\cal L}}
\def\c{c_\chi}
\def\s{s_\chi}
\def\xh{\hat x}
\def\yh{\hat y}
\def\zh{\hat z}
\def\vt{\widetilde{v}}
\def\wt{\widetilde{w}}
\def\n#1{{\hat{n}_{#1}}}
\def\c#1{\cos{#1}}
\def\s#1{\sin{#1}}
\def\cs#1{\cos^2{#1}}
\def\ss#1{\sin^2{#1}}

\title{Dirichlet Solitons in Field Theories}

\author{Mark Trodden\footnote{{\tt trodden@theory1.physics.cwru.edu.} \\
Talk presented at Cosmo-98, Particle Physics and the Early Universe,
Asilomar, CA November 15-20, 1998}}

\address{~\\Particle Astrophysics Theory Group,
Department of Physics,
Case Western Reserve University,
10900 Euclid Avenue,
Cleveland, OH 44106-7079, USA.}

\maketitle

\begin{abstract}
I briefly describe a new class of soliton configurations in field theories. 
These consist of topological defects which can end when they intersect other 
defects of equal or higher dimensionality.  Such configurations may be termed
``Dirichlet topological defects'', in analogy with the D-branes of
string theory.  I provide a specific example - cosmic strings that terminate
on domain walls - and discuss some new directions for this work, including
an interesting and qualitatively different extension to supersymmetric 
theories.
\end{abstract}

\vspace{5mm}
CWRU-P2-99
\vspace{8mm}

The study of topological soliton solutions to classical field theories has
led to many important new ideas in particle physics and cosmology. In particle 
physics, a common feature of the recent progress in both supersymmetry (SUSY)
and string theory has been the discovery of {\it dualities}. These dualities
map the calculationally difficult limit of one theory into the (hopefully)
calculationally easier limit of another theory. More precisely, dualities
often interchange the roles of the fundamental and solitonic degrees of 
freedom. For the case of Seiberg-Witten dualities in SUSY field theories, the 
relevant solitons are monopoles, so that the electric and magnetic degrees
of freedom are interchanged. For string theories, the relevant objects are
the D-branes; extended configurations on which fundamental strings can
end. In both cases, the study of solitons in the theories has led to a better 
understanding of how particle physics works.

In cosmology, solitons, or {\it defects}, play two main roles. First, one
may use cosmology to constrain candidate particle physics theories. Examples
of this include the requirement that symmetry breaking schemes not allow
magnetic monopoles in the early universe, and vorton constraints on theories
admitting superconducting cosmic strings. Second, the dynamics, interactions, 
and microphysics of topological defects can provide explanations for
cosmological problems. Examples of this include the use of cosmic strings and
textures as seeds for the large scale structure, and the idea that
topological solitons might play a role in the generation of the baryon 
asymmetry of the universe.

Topological defects are solitonic solutions whose stability is guaranteed 
by a topological conservation law. 
When a symmetry group $G$ is spontaneously broken to a subgroup $H$, the 
types of defects supported depend on the homotopy properties
of the vacuum manifold, $\M = G/H$.
In a $(d+1)$-dimensional spacetime, $p$-dimensional defects ($p<d$)
exist if the homotopy group $\p{d-p-1}(\M)$ is nontrivial. 
(For reviews see \cite{reviews}.)
In addition to these basic defects, there are various
composite solutions which combine two of the types, generally when
a $(p-1)$-dimensional defect serves as the boundary of a $p$-dimensional
defect. Such configurations have interesting cosmological applications, such 
as in the Langacker-Pi mechanism for solving the monopole problem \cite{LP}.

In this brief review I will describe a new class of topological defects 
\cite{dirichlet}
which is complementary to those mentioned above.
These consist of field configurations in which one type of topological 
defect can terminate when intersecting other defects of equal or higher 
dimensionality.  Such configurations may be termed ``Dirichlet topological 
defects'', in analogy with the D-branes of string theory.  The latter
are extended objects on which fundamental strings can end
\cite{horava}.  The models considered here are ordinary 
field theories, which support topological solitons which resemble these
objects in fundamental string theory.  Here I will only have space to
provide the specific example of cosmic strings that terminate on domain 
walls, and to comment on future directions for investigation.

Strings arise most simply from the breakdown of $\u1$
symmetries. Therefore consider two complex fields 
$\psi_i = \rho_i e^{i\theta_i}$, ($i=1,2$), and a single real scalar $\phi$,
transforming under two $\u1$ and one $\Z_2$ symmetries in the following way

\begin{eqnarray}
 \Z_2:&\ \ \ &\{\phi \rightarrow -\phi\ ,\ \psi_1 \leftrightarrow \psi_2\} \ ,
 \nonumber\\
 {\u1}_1:&\ \ \ & \psi_1  \rightarrow e^{-i\omega_1}\psi_1\ , 
 \label{symm2}\\
 {\u1}_2:&\ \ \ & \psi_2  \rightarrow e^{-i\omega_2}\psi_2\ .
  \nonumber
\end{eqnarray}
The two $\u1$'s may be taken to be either global or gauge symmetries. 
In the latter case,
$\omega_1$ and $\omega_2$ are functions of spacetime, and
there are two gauge fields $A_\mu^{(1)}$, $A_\mu^{(2)}$, with
the usual transformation properties, and associated covariant derivatives.

Write a general, renormalizable potential in the convenient form

\begin{eqnarray} 
  V(\phi,\psi_1,\psi_2) 
  & = & \lambda_\phi(\phi^2 - \vt^2)^2 + \lambda_\psi\left[|\psi_1|^2 + 
  |\psi_2|^2
   - \wt^2+ g(\phi^2  - \vt^2)\right]^2
  \nonumber\\
  & & \quad {} + h |\psi_1|^2 |\psi_2|^2 - \mu\phi(|\psi_1|^2 - |\psi_2|^2)\ ,
  \label{pot2}
\end{eqnarray}
where  $v=\langle |\phi |\rangle$ is the root of the cubic equation

\begin{equation} 
  8\lambda_\phi\lambda_\psi v^3 + 6\lambda_\psi g\mu v^2 
  - (8\lambda_\phi\lambda_\psi\vt^2 + \mu^2)v 
  - 2\lambda_\psi (g\vt^2 + \wt^2)\mu  =0 
\end{equation} 
that reduces to $\vt$ at $\mu = 0$, and $w$ is given by
\begin{equation} 
  w = \left(\wt^2 + g(\vt^2 - v^2) +
  {\mu v \over{2\lambda_\psi}}\right)^{1/2}
  \ .
  \label{vev2}
\end{equation} 
In the vacuum the real scalar $\phi$ takes the vev $\pm v$
and there may exist domain walls separating these two values.  When
$\langle\phi\rangle=+v$, the vacuum has $|\psi_1| = v$ and $\psi_2 = 0$,
while when $\langle\phi\rangle=-v$ the vacuum has $|\psi_2| = v$ and  
$\psi_1 = 0$.  

In this model, therefore, the unbroken symmetry group in the true
vacuum is $\u1$,
and the vacuum manifold is $\M = [\u1 \times\u1 \times \Z_2]/\u1
= S^1 \times \Z_2$, admitting walls and strings.  
When $\langle\phi\rangle=+v$, the complex field
$\psi_1$ can form cosmic strings with winding number $n$, around
which $\theta_1$ will change by $2\pi n$.  Such a string ends if
it intersects a D-wall, since $\langle\psi_1\rangle=0$ on the other 
side.  Analogous statements hold for the $\psi_2$ field when
$\langle\phi\rangle=-v$.

In the core of a string the
corresponding $\u1$ symmetry is restored.  In the gauge case,
therefore, the gauge bosons associated with, for example, ${\u1}_1$ 
are massless both in the core of a $\psi_1$-string on the
$\langle\phi\rangle = v$ side of the D-wall, and anywhere on the
$\langle\phi\rangle = -v$ side of the D-wall.  As usual, outside
the $\psi_1$-string the gauge field is pure gauge,
such that it cancels the gradient energy of the scalars by  
enforcing the vanishing of the covariant derivative.  
The gauge field is thus given by
$A_\mu^{(1)}  =  - \partial_\mu\theta_1$.
Consequently, there is magnetic flux through the string (which
I take to have winding number $n$), given by $\Phi^{(1)} = -n\pi$.
This flux flows through the string until it hits the wall; on the
other side of the wall the symmetry is unbroken everywhere, and the
magnetic field describes a monopole configuration emanating from
the point where the string intersects the wall.

Configurations of the this type, with strings ending on walls, have
recently been discussed in the context of supersymmetric QCD \cite{qcd}.
There, the string consists of non-Abelian flux, and the wall
separates different chiral vacua, with shifted values of the QCD
$\theta$-parameter.  The intersections of strings and domain walls
can be thought of as quarks.  
The structures of these QCD configurations
and the scalar field models discussed here are obviously very similar,
and the relationship between them deserves further investigation.
(One difference is that the flux in the strings considered in \cite{qcd}
does not propagate freely on the other side of the wall, as the symmetry
is still broken there; rather, it is confined to the wall itself.
It should not be difficult to extend models of the type considered in
this paper to include such situations.) 

I have described a class of topological defects in
classical field theories in (3+1) dimensions, consisting of Dirichlet
defects on which fundamental defects of lower dimension can
terminate.  While the search for models supporting
these configurations is inspired by the appearance of D-branes in
string theory, there are important differences between the two
sets of objects.  In all of the theories considered, the
basic degrees of freedom are scalar fields and gauge fields,
out of which all of the higher-dimensional objects are constructed.
Gravity and supersymmetry are not included (although there are no 
obstacles to the appropriate generalizations \cite{susydefects} ).  
Furthermore, the specific dependence of D-brane
energy on the string coupling constant is not a feature of our models,
and the Ramond-Ramond gauge fields to which D-branes couple are
absent.  Nevertheless, it may be interesting to compare the
dynamical behavior of Dirichlet defects to that of D-branes in
string theory, and search for models in which the similarities 
between the two systems are even stronger.

One obvious direction in which to generalize the models considered
here is to consider $q$-dimensional defects ending on
$p$-dimensional D-defects in $d$ spatial dimensions.  (There are
a variety of such objects in string theory and M-theory, with
configurations governed by charge conservation.)
A number of
interesting issues arise in this case, especially for gauge
symmetries. For example, to make
topological defects of dimension $q$ in $d$ spatial dimensions
requires that $\p{d-q-1}(\M)$ be nontrivial, for example by
breaking $\so{d-p}$ to $\so{d-p-1}$ (for which $\M = S^{d-p-1}$).
In such a model, the unbroken symmetry group $\so{d-p-1}$ is
non-Abelian for $p\leq d-4$ and the low-energy gauge theory is expected
to be strongly coupled, and the resulting defects to be confined.

Finally, as with any species of topological defect,
it is also natural to ask what the cosmological consequences of
the formation of these objects in the early universe might be.

\section*{Acknowledgments}
I would like to thanks my collaborators, Sean Carroll and Simeon Hellerman
for a lot of fun working on these topics. Thanks also to the organizers, for
a great meeting. This work was supported by the Department of
Energy (D.O.E.).

\end{document}